# Automatic Extraction of Agriculture Terms from Domain Text: A Survey of Tools and Techniques

**Niladri Chatterjee, Neha Kaushik**

Niladri Chatterjee and Neha Kaushik are with the Indian Institute of Technology Delhi, Hauz Khas, New Delhi-110016, India, phone: 011-26591490;  e-mail: niladri@maths.iitd.ac.in, neha@kitd.ac.in

**ABSTRACT**

Agriculture is a key component in any country's development. Domain-specific knowledge resources serve to gain insight into the domain. Existing knowledge resources (AGROVOC and NAL Thesaurus) are developed and maintained by domain experts. Population of terms into these knowledge resources can be automated by using automatic term extraction tools for processing unstructured agricultural text. Automatic Term Extraction is also a key component in many semantic web applications, such as ontology creation, recommendation systems, sentiment classification, query expansion among others. The primary goal of an automatic term extraction system is to identify important terms of a specific domain from a given set of text documents. The objective of such a system is to maximize the number of valid terms and minimize the number of invalid terms extracted from the input set of documents. Despite its importance in various applications, availability of online tools for the said purpose is rather limited. Moreover, the performance of the most popular ones among them varies significantly. As a consequence, selection of the right term extraction tool is perceived as a serious problem for different knowledge-based applications.  This paper presents an analysis of three commonly used term extraction tools, viz. RAKE, TerMine, TermRaider, and compares their performance in terms of Precision and Recall, vis-à-vis  RENT, a more recent term extractor developed by these authors for agriculture domain.

1. **INTRODUCTION**

Agriculture plays an important role in the growth of any country. Proliferation of digitization everywhere leads to exponential growth in creation of electronic data of a domain. Domain-specific terms play an important role in understanding the domain, their roles and significance, and terminologies, and also the relationship between them.

Agriculture data in the form of unstructured text can be processed to extract important terms of the domain automatically. This will aid in automatic creation of agriculture-specific knowledge resources. Two existing agriculture-specific knowledge resources are AGROVOC[1] and NAL thesaurus[2]. These are developed and maintained by human experts.

The automatically extracted terms can also be used for further expansion of existing resources. Domain experts can select the significant terms from the collection of terms extracted using automatic term extraction tools.

In this paper, we provide an insight into automatic term extraction techniques and  present a comparison of four automatic term extraction tools with reference to agriculture domain.

2. **AUTOMATIC TERM EXTRACTION TECHNIQUES**

Automatic Term Extraction, with the help of Statistical, Distributional, Contextual and Linguistic methods [1], aim to identify the important terms pertaining to a specific domain from large electronic text data. Automatic Term Extraction may be considered as a first step in many semantic web applications, such as, sentiment classification [2], ontology creation [1,3], query expansion [4], automated knowledge provider systems [5]   and recommendation systems [6].

Basic techniques for automatic term extraction can be classified into four categories:

---

[1] http://aims.fao.org/vest-registry/vocabularies/agrovoc
[2] https://agclass.nal.usda.gov/

*Statistical Methods*

These methods employ various statistical measures of words in a corpus to identify important terms in a domain. Different statistical measures, such as raw frequency count [7], Point-wise mutual information (PMI) [8], Dice coefficient [8], log likelihood ratio [8] have been used in literature for extracting terms.

*Distributional Methods*

Distributional methods [9] are specialized form of statistical methods which use distribution of words in the corpus besides their frequencies. Examples of distributional measures are term frequency and inverse document frequency (TF-IDF) [10] and weirdness [11].

*Contextual Methods*

Contextual methods are based on the assumption that context of a word plays an important role in identifying the meaning of that word. In such methods, internal structure and context of words, besides their frequencies, are used for extraction of terms from the text. NC-Value [12] method is an example of contextual methods. [13] presents multi word term extraction using contextual information.

*Linguistic Methods*

Linguistic methods identify terms based upon their syntactic properties [14]. Typical sequence of operations in such methods is part-of-speech tagging (POS tagging), stop words removal and language filters.
POS tagging is assignment of part of speech to each word in the text.
Extremely frequent, very generic words, commonly known as *stop words*, which rarely form a valid term, are often removed from the text as a pre-processing step to automatic term extraction algorithms, leading to improved precision, in general.
Basic concept of linguistic methods is that nouns and adjectives are the most appropriate candidates for being important terms of a domain. Language filters [12] can be used to identify phrases fulfilling specific forms, such as NP (Noun Phrase), Proposition Phrase (PP).
Often, a hybrid of these techniques is used for automatic term extraction. Hybrid methods [15] use a combination of two or more types of methods discussed above.
The analysis presented in this paper is based on agriculture as the underlying domain of interest. The domain is so chosen because of the vastness of data and under-utilization of data processing methods to the domain. Agriculture domain is composed of various sub-domains namely fertilizers, irrigation, crops, regions, diseases, seeds, farming systems, soil, weather, pesticides among others [16]. We have worked and analyzed the results obtained from the four automatic term extraction tools mentioned earlier with respect to identification of terms in the domain.

The objective of the present paper is to provide the state-of-the-art techniques in the field of automatic term extraction, and to compare three well-known Automatic Term Extraction tools, viz, RAKE [17], TerMine [12], TermRaider [18] vis-à-vis RENT [19], one of the latest term extraction scheme proposed in recent time. The paper further focuses on identifying the challenges and future directions to enhance the study. The paper is organized as follows : Section 2 presents an overview of the four automatic term extraction tools analyzed in this research, Section 3 presents the experimental setup for the analysis, Results and Conclusion are presented in Section 4 and Section 5 respectively.

## 3. AUTOMATIC TERM EXTRACTION TOOLS

### 3.1 RAKE

Rapid Automatic Keyword Extraction (RAKE) [17] is a simple tool for automatic term extraction. Figure 1 shows the flow of algorithm for RAKE, explained in the following paragraphs.
The first step in RAKE algorithm is document parsing. It consists of extracting the phrases delimited by stop words and punctuation. Text documents, list of stop words, phrase delimiters (punctuation marks, such as fullstop (.), comma (,), apostrophe (')) are the inputs given to the algorithm at this step. Document parsing and hence candidate keyword extraction in this manner enables the identification of word co-occurrences without the need of an arbitrarily sized sliding window.

Second step is computation of keyword scores. The score of a candidate keyword is sum of the scores of its constituent words. Score of each constituent word is calculated as

$$score(w) = \frac{degree(w)}{frequency(w)}$$

where
- *degree* ($w$) = measure of the co-occurrence of $w$ with other words in a keyword and length of the keywords in which $w$ occurs
- *frequency*($w$) = number of times $w$ occurs in the document
- *score*($w$) ranks consistently co-occurring words higher than a word which is highly frequent but is dispersed randomly through the document.

The third step in RAKE algorithm focuses on keywords containing *interior stop words* (e.g. off season cultivation, here off is a stop word which is part of a keyword. RAKE identifies adjoining keywords which occur at least three times together in the same document and in the same order. New candidate keywords are appended to the list of candidate keywords by including a combination of such adjoining keywords and their interior stop words. This step is useful for large texts only, as it looks for multiple occurrences of adjoining keywords within the same document.

The fourth and final step of RAKE consists of identifying the actual list of keywords from the set of candidate keywords. This is done by taking out the top $T$ terms from the set of candidate keywords. Often, $T$ is taken to be one third of the total number of candidate keywords identified by RAKE in the previous steps.

RAKE is available as a library named rake-nltk[3] in Python. Given an input text, it outputs the keywords with their scores. However, in our experiments we observed that it does not perform well for domain-specific purpose and extracts many irrelevant and lengthy terms.

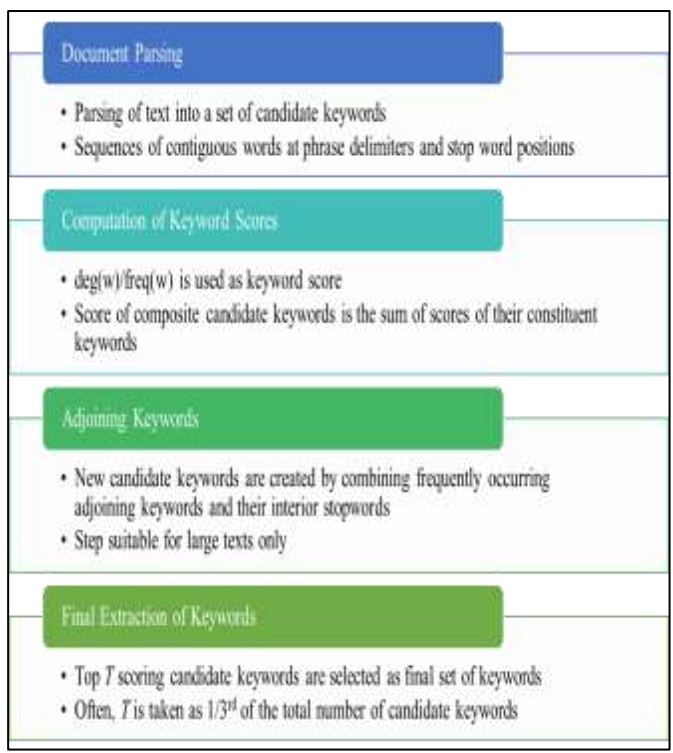

Figure 1. Overview of RAKE

---

[3] https://pypi.org/project/rake-nltk/

## 3.2 TerMine

TerMine [12] is a hybrid method for term extraction based on C-Value method. The equation for C-Value method is:

$$C-value(t) = \begin{cases} \log_2|t| \cdot f(t) & t \text{ is not nested} \\ \log_2|t|\left(f(t) - \frac{1}{P(N_t)}\sum_{v \in N_t} f(v)\right) & \text{otherwise} \end{cases}$$

where
- $t$ is the candidate string,
- $f(t)$ is the frequency of occurrence of $t$,
- $N_t$ is the set of extracted candidate terms that contain $t$,
- $P(N_t)$ is the number of these terms,
- $v$ belongs to the set $N_t$ and
- $f(v)$ is the frequency of occurrence of $v$.

Figure 2 shows an overview of the component analysis used by TerMine for automatic term extraction.
This method uses a combination of linguistic and statistical information for the said purpose. Linguistic information includes POS tagging of the corpus, the linguistic filter, and the stop list. Statistical part includes the total frequency of occurrence of the candidate string in the corpus, frequency of candidate string as part of other longer candidate terms (nested string), the number of such nested terms, length of the candidate string (in number of words).
C-value method used in this tool is often treated as a benchmark in Term Extraction [20]. The tool performs much better compared with RAKE. However, one limitation of this tool is that it extracts only multi-word terms. This is major shortcoming as in a domain where many important single word terms often constitute the set of important terms.

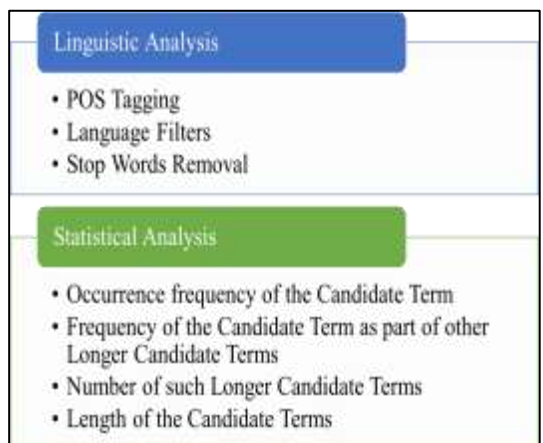
Figure 2. Overview of TerMine

## 3.3 TermRaider

TermRaider[4] [18] is available as a plug-in with GATE (General Architecture for Text Engineering)[5]. It requires a corpus consisting of domain related documents for term extraction. The tool makes use of Linguistic and distributional methods for the aforesaid purpose. Here, the input corpus is pre-processed using linguistic methods such as POS tagging and stop words removal. ANNIE plug-in is loaded for this purpose in GATE. The TermRaider plug-in extracts the useful terms from the corpus based upon TF-IDF score.
An overview of the term extraction process using TermRaider is presented in Figure 3.Tokenization and sentence splitting divides the input text into manageable units. These units are further processed using lemmatization and POS tagging. This helps include the internal structure of words and their organization into phrases (collectively called as morpho-syntax) into the analysis. Candidate terms are then filtered out using language filters to identify noun phrases. TF-IDF is used to compute the score of each candidate term. All terms having score more than a manually selected threshold are considered to be final list of valid terms.

---
[4]https://gate.ac.uk/projects/neon/termraider.html
[5]https://gate.ac.uk/

TF and IDF stand for Term Frequency and Inverse Document Frequency, respectively. TF-IDF [10] measures the importance of a term to a document in a given corpus. IDF is the inverse of the proportion of documents where the term $t_i$ occurs, scaled to logarithmic scale as given below:

$$idf(w_i) = log\left(\frac{D}{D_i}\right)$$

Here, $D$ is the total number of documents in a collection, and $D_i$ the number of documents in which $w_i$ occurs. Combining term frequency and inverse document frequency the term weight (*tf-idf*) of a word for a document is defined as the product *tf * idf*. Hence,

$$tf\text{-}idf(w_i) = f(w_i) * log\left(\frac{D}{D_i}\right)$$

The hypothesis behind *tf-idf* is that important terms have high frequency in the domain but these tend to be distributed in few number of documents of the domain.

TermRaider needs a lot of documents for good performance. It cannot be used if one wants to extract useful terms from a single domain-specific document. A corpus is necessarily required for extracting terms using TermRaider.

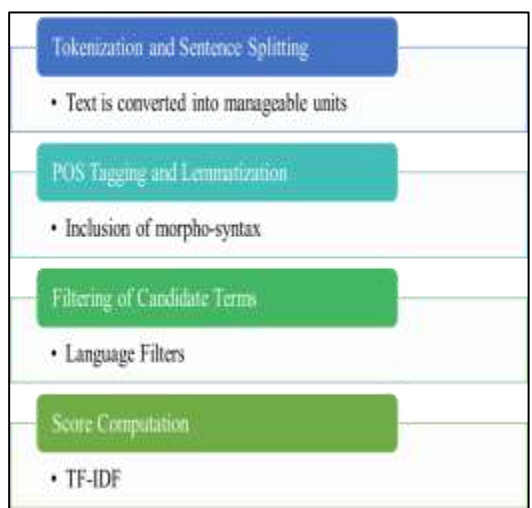

Figure 3. Overview of TermRaider

## 3.4 RENT

*R*egular *E*xpression and *N*LP-based *T*erm Extraction Scheme [19] uses domain-specific patterns identified by experts, encoded in the form of regular expressions to identify seed terms. Stop words are removed from the set of seed terms in the second step. Further terms are extracted from the text using linguistic filters. Combination of noun with other nouns, adjective and noun is considered to be a valid term if either of the constituent word is present in the set of seed terms. Computation of weights for each terms is done using three components:
- POS information, weight of a word, *weight*($w_i$), is incremented by 1 if it is a noun
- No. of regular expressions that a word satisfies, for each pattern with which a term occurs, *weight*($w_i$) is incremented by 1
- Frequency of occurrence, finally *weight*($w_i$) is updated as follows:

$$weight(w_i) = a * weight(w_i) + b * frequency(w_i)$$

where *frequency*($w_i$) is the frequency of occurrence of $w_i$.

Figure 4 provides an overview of RENT algorithm. The RENT algorithm, although designed for agriculture domain, can suitably be modified for other domains with minor changes in the regular expressions used to identify the set of seed terms.

## 4. EXPERIMENTAL SETUP AND RESULTS

Agriculture related text (input text) is collected from Department of Agriculture's website using the links *farmer.gov.in*, *agricoop.nic.in*, handbooks available from FAO, *nios.ac.in* etc. Ten samples of text are taken in the form of 5 pages each from the input text.

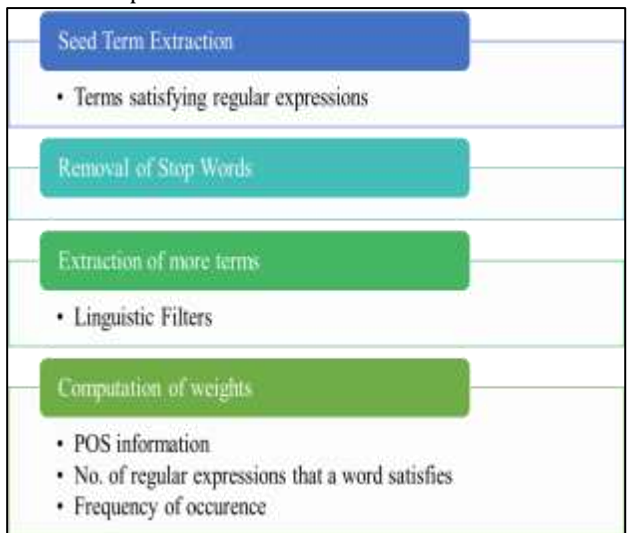
Figure 4. Overview of RENT

Automatic term extraction tools presented in Section 2 are run on these samples in a cumulative manner, i.e. in the $k^{th}$ iteration, the input text consists of $5*k$ pages.

Precision and recall values for $i^{th}$ iteration are computed using the following formulae:

$$p_i = \frac{Nv_i}{Nt_i}$$

$$r_i = \frac{Nvi}{T_i}$$

where
- $p_i$ = precision for the $i^{th}$ iteration of the execution,
- $r_i$ = recall for the $i^{th}$ iteration of the execution,
- $Nv_i$ = number of valid terms extracted from first $i$ sets of 5 pages,
- $Nt_i$ = total number of terms extracted using the corresponding tool from the first $i$ sets of 5 pages,
- $T_i$ = total number of terms identified by experts to be present in the first $i$ sets of 5 pages.

Table 1 presents the precision and recall values for each iteration of execution of these four tools. The average values of precision and recall for each tool are shown in bold.

It can be seen from Table 1 that RENT algorithm outperforms TermRaider, TerMine and RAKE in terms of precision on an average (as well as for every iteration). This is attributed to the regular expressions and the re-sequencing of text processing steps in the term extraction process. RENT algorithm also outperforms in terms of average recall. However, it can be seen that RAKE algorithm performs better than other algorithms in terms of recall in the last four iterations. This means that RAKE performs well in terms of recall as the input size increases. TerMine performs good in terms of precision, it lacks in recall because it is designed to extract only the multiword terms.

## 5. CONCLUSION

This paper provides an overview of Automatic Term Extraction. It compares four popular tools for automatic term extraction in terms of precision and recall.

RENT algorithm performs efficiently in terms of average precision and recall. It has an average precision of 80% approximately. The average value of recall is 45.11.

RENT algorithm can be modified so as to include more number of valid terms for improvement of recall. This can be achieved by increasing the number of regular expressions after careful analysis of the domain. Recall of RENT algorithm can also be improved by inclusion of terms from domain-specific lexical resources, related to the seed terms identified in RENT algorithm. Examples of such lexical resources for agriculture domain include AGROVOC, NAL thesaurus and WordNet.

RENT algorithm can be applied to other domains as well. This can be done by adapting the regular expressions in accordance with the domain.

Automatic term extraction is a significant task in another knowledge representation technique, viz. ontology development. Terms identified using automatic term extraction serve as good input for creation of concepts in an automatic ontology development environment.

**Table 1. Precision and Recall for TerMine, RENT, TermRaider and RAKE**

| No. of pages | TerMine | | RENT | | TermRaider | | RAKE | |
|---|---|---|---|---|---|---|---|---|
| | Precision % | Recall % | Precision % | Recall % | Precision % | Recall % | Precision % | Recall % |
| 5 | 50.00 | 26.62 | 80.76 | 45.56 | 48.78 | 30.17 | 58.82 | 36.09 |
| 10 | 54.03 | 17.13 | 72.30 | 54.49 | 44.72 | 25.28 | 50.85 | 28.09 |
| 15 | 47.91 | 12.77 | 81.73 | 52.57 | 45.70 | 30.88 | 56.40 | 40.19 |
| 20 | 55.93 | 26.22 | 78.61 | 41.26 | 41.32 | 26.02 | 52.34 | 36.27 |
| 25 | 50.60 | 23.52 | 82.00 | 42.54 | 39.47 | 27.25 | 53.73 | 41.96 |
| 30 | 53.58 | 21.72 | 85.81 | 44.88 | 39.11 | 28.18 | 52.62 | 43.44 |
| 35 | 57.24 | 20.73 | 79.64 | 42.73 | 39.65 | 30.21 | 54.18 | 48.41 |
| 40 | 60.44 | 21.19 | 77.07 | 42.05 | 38.19 | 30.84 | 51.34 | 43.11 |
| 45 | 56.81 | 19.87 | 83.57 | 40.06 | 37.53 | 32.96 | 47.47 | 44.70 |
| 50 | 57.02 | 25.41 | 77.75 | 44.96 | 36.36 | 31.72 | 47.18 | 45.03 |
| **Average** | **54.35** | **21.51** | **79.92** | **45.11** | **41.08** | **29.35** | **52.49** | **40.72** |

## References

1. D. Maynard, Y. Li and W. Peters, "NLP Techniques for Term Extraction and Ontology Population", In Proceedings of the 2008 Conference on Ontology Learning and Population: Bridging the Gap between Text and Knowledge, pp. 107-127.
2. T. Nasukawa and J. Yi, "Sentiment Analysis: Capturing Favoribility using Natural Language Processing", In Proceedings of the 2nd International Conference on Knowledge Capture, ACM, 2003, pp. 70-77.
3. P. Velardi, P. Fabriani and M. Missikoff, "Using Text Processing Techniques to Automatically Enrich a Domain Ontology", In Proceedings of the International Conference on Formal Ontology in Information Systems – Volume 2001, ACM, 2001, pp. 270-284.
4. S.G. Shaila and A. Vadivel, "TAG Term Weight-based N gram Thesaurus Generation for Query Expansion in Information Retrieval Application" in *Journal of Information Science*, Vol. 41 no. 4, pp. 467-485, 2015.
5. P. Mukherjee and B. Chakraborty, "Automated knowledge provider system with natural language query processing," *IETE Tech. Rev.*, Vol. 33, no. 5, pp. 525–38, Dec. 2015.
6. A. Collins and J. Beel, "Document Embeddings vs. Keyphrases vs. Terms for Recommender Systems: A Large-Scale Online Evaluation" InACM/IEEE Joint Conference on Digital Libraries (JCDL), Champaign, IL, USA, 2019, pp. 130-133.
7. W.B. Frakes, G. Kulczycki, J. Tilley, "A Comparison of Methods for Automatic Term Extraction for Domain Analysis", In Schaefer I., Stamelos I. (eds) Software Reuse for Dynamic Systems in the Cloud and Beyond. ICSR 2015. Lecture Notes in Computer Science, Springer, Cham, Vol. 8919, pp. 269-281, 2015.
8. L. Ahrenberg, "Term Extraction: A Review", Draft Version 091221, 2009. Available at, http://www.ida.liu.se/~larah03/Publications/tereview_v2.pdf
9. R. Nazar, "Distributional Analysis applied to Terminology Extraction", In *Terminology*, Vol. 22, No. 2, pp. 141-170, 2016.
10. Z. Zhang, J. Iria, C. Brewster, and F. Ciravegna, "A comparative evaluation of term recognition algorithms", In Proceedings of the Sixth International Conference of Language Resources and Evaluation (LREC-08), pp. 2108-2111, 2008.
11. K. Ahmad, L. Gillam, and L. Tostevin, "Weirdness Indexing for Logical Document Extrapolation and Retrieval (WILDER)", In The Eighth Text Retrieval Conference (TREC-8), 1999.
12. K. Frantzi, S. Ananiadou, H. Mima, "Automatic Recognition of Multi-word Terms: The C-Value/NC-Value Method", In *International Journal on Digital Libraries*, Vol. 3, No. 2, pp. 115-130, 2000.
13. D. Maynard and S. Ananiadou, "Identifying contextual information for Multi-word Term Extraction", In proceedings of Terminology and Knowledge Engineering Conference, Vol. 99, pp. 212-221, 1999.
14. M. T. Pazienza, M. Pennacchiotti, and F. M. Zanzotto, " Terminology Extraction: An Analysis of Linguistic and Statistical Approaches", In *Knowledge Mining*, pp. 255-279, 2005.
15. da S. Conrado, A. D. Felippo, T.A. S. Pardo, and S.O. Rezende, "A Survey of Automatic Term Extraction for Brazilian Portuguese", In *Journal of Brazilian Computer Society*, Vol. 2, No.1, pp. 12-39, 2014.
16. N. Chatterjee, N. Kaushik and B. Bansal, "Inter-Subdomain Relation Extraction for Agriculture Domain", In *IETE Technical Review*, vol. 36 no. 2, pp. 157-163, March, 2018.
17. S. Rose, D. Engel, N. Cramer and W. Cowley, "Automatic Keyword Extraction from Individual Documents", In Text Mining: Applications and Theory, pp. 1-20, 2010.
18. D. Maynard, Y. Li, and W. Peters, "NLP techniques for term extraction and ontology population", In Proceedings of the 2008 Conference on Ontology Learning and Population: Bridging the Gap Between Text and Knowledge, pages 107–127, Amsterdam, The Netherlands, The Netherlands. IOS Press,2008.
19. N. Kaushik and N. Chatterjee, "RENT: Regular Expression and NLP-Based Term Extraction Scheme for Agricultural Domain", In Proceedings of the International Conference on Data Engineering and Communication Technology, India, 2017, pp. 511-522.
20. R. Wang, W. Liu, & McDonald C.: "Featureless Domain-Specific Term Extraction with Minimal Labelled Data", In Proceedings of the Austrlian Language Technology Association Workshop, 2016, pp. 103-112, 2016.